# Quasi-steady-state air waveguide

## A. Goffin, A. Tartaro, and H.M. Milchberg[*]

*Institute for Research in Electronics and Applied Physics, University of Maryland, College Park, MD, 20742, USA*
**milch@umd.edu*

**We report the first generation of quasi-steady-state air waveguides capable of guiding high average power laser beams. The guides are produced by high-repetition rate patterned filamentation of femtosecond laser pulses. We demonstrate near-continuous guiding of a CW probe beam with significantly higher efficiency than transient guides at lower repetition rates.**

Femtosecond laser filaments [1,2] have been used for a variety of applications, including guiding of high voltage discharges [3,4], fog clearing [5,6], light detection and ranging (LIDAR) [7], laser-induced breakdown spectroscopy (LIBS) [8], and air lasing [9]. One application of note is femtosecond filament generation of long-lasting air waveguides capable of guiding much higher average power beams than the filaments themselves [10]. Most recently, we demonstrated air waveguiding over ~50 m, with guide lifetimes of ~20 ms limited by thermal diffusion of the imprinted waveguide cladding [11]. A path to a quasi-continuous air waveguiding was thus provided by [10,11], along with the necessary understanding of gas dynamics induced by high-repetition-rate filamentation [12,13]. In this paper, we demonstrate, for the first time, quasi-continuous air waveguiding using kHz repetition rate filamentation for generating and sustaining the waveguide.

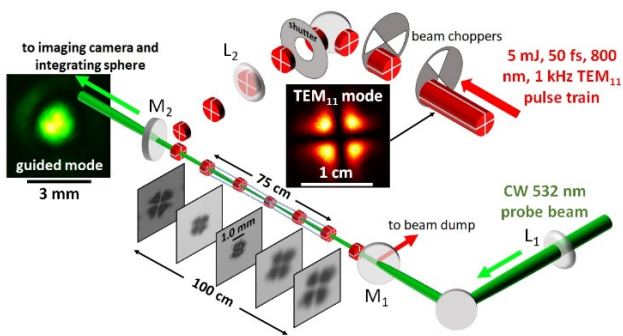

**Figure 1.** Air waveguiding experiment. $M_1/M_2$: dielectric mirrors to reflect 800nm and transmit 532nm. $L_1$: lens for f/200 injection of CW probe into guide. $L_2$: lens for f/250 focusing of the filamenting $TEM_{11}$ beam.

The experimental setup is depicted in Fig. 1. The air waveguide is generated at repetition rates 10-1000 Hz by filamentation of Ti:sapphire laser pulses in a (4-lobed) $TEM_{11}$ mode, produced by passing a near-Gaussian beam through a four-segment binary phase plate, followed by a spatial filter. The pulses are focused by lens $L_1$ in air at f/250, forming ~75 cm long filaments, after which the $\lambda = 800$ nm light is removed from the beam path by dielectric mirror M1. The filament repetition rate is controlled by two chopper wheels in the path of the 1 kHz repetition rate laser, and a millisecond risetime shutter is used for burst mode experiments. Each lobe of the $TEM_{11}$ mode is $\sim 3P_{cr}$ (critical power for self-focusing) in air [14]. Burn paper was used to monitor filamentation of the $TEM_{11}$ pulse over its propagating path, with burn images shown in Fig. 1. The diagonal filament separation near the beam waist is ~500 μm, so that the waveguide core diameter is ~200 μm after ~0.5 ms of thermal diffusion and merging of the four filament-induced density holes to form the guide cladding [10].

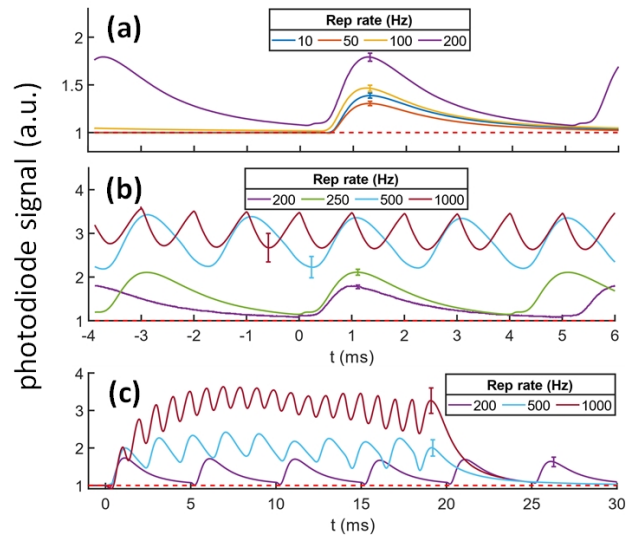

**Figure 2.** Integrating sphere photodiode traces of air-waveguided probe beam. All curves are 200 trace averages. Error bars shown are the standard deviation over individual traces. The dashed red line is the waveguide-off photodiode signal, normalized to unity. **(a)** Repetition rates 10-200 Hz **(b)** Transition to quasi-continuous operation over repetition rates 200-1000 Hz. **(c)** Probe signal showing air waveguide buildup from $t = 0$ in burst mode at 200 Hz, 500 Hz, and 1000 Hz.

Optical guiding is measured by injecting at $f/200$ a counter-propagating, spatially filtered continuous wave laser beam ($\lambda = 532$ nm, 4.5 mW) into the air waveguide. This beam is then either imaged 50 cm after the waveguide exit onto a CCD camera or directed into an integrating sphere with a photodiode. The integrating sphere entrance is apertured with an adjustable iris to maximize the ratio of guided (waveguide on) to unguided (waveguide off) beam signal at each filament repetition rate. This ensures that the waveguide-on signal originates from the waveguide core.

The effect of filament repetition rate is shown in Fig. 2. For 10 Hz through 100 Hz, the guided signal decays over ~4 ms from its peaks to the waveguide-off level (red dashed line) before the next TEM$_{11}$ pulse arrives. This decay time for ~200 µm core diameter guides is consistent with our previous measurements and calculations [10-12] of the decline of the core-cladding index contrast by thermal diffusion. Beyond 200 Hz, the guided beam signal is always higher than the waveguide-off level, indicating the transition to a quasi-continuous air waveguide. As seen in Fig. 2(b), improved guiding is observed at higher repetition rates, with a large increase in guided signal between 200 Hz and 500 Hz. While the peak values of the 1 kHz signal are similar to those at 500 Hz, the dips between peaks are reduced at 1 kHz as the guide approaches steady state operation. The increase in guide efficiency is driven by the cumulative deepening of filament-induced density holes at high repetition rate, as first measured in ref. [12], resulting in a deeper cladding and a higher core-cladding index contrast. The effect of an abrupt turn-on at $t = 0$ of the waveguide-generating pulse train is shown in Fig. 2(c), where a ~ 1 ms risetime shutter is placed in the TEM$_{11}$ beam. The rapid cumulative growth of waveguide efficiency is clearly demonstrated by the >200 Hz traces.

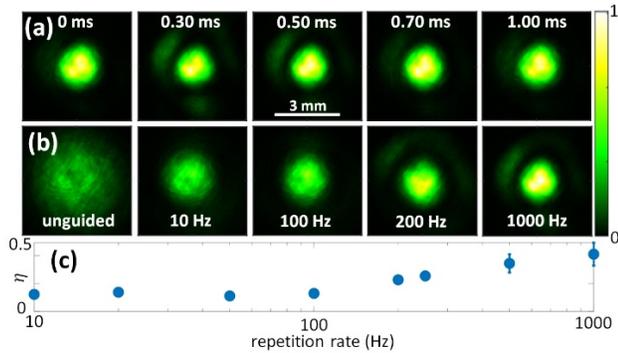

**Figure 3. (a)** Guided mode evolution in the interval between TEM$_{11}$ pulses for a 1 kHz waveguide. **(b)** Guided mode vs. repetition rate at 500 µs delay. **(c)** Guide efficiency $\eta$ vs. repetition rate at 500 µs delay.

Figure 3(a) shows guided mode evolution for a 1 kHz waveguide during the interval between TEM$_{11}$ pulses (here $t = 0$ corresponds to a peak in the 1 kHz trace in Fig. 2(b)). CCD camera images are taken using a 20 µs electronic shutter, much shorter than the thermal evolution of the waveguide, so the contribution of unguided light is negligible. The probe beam is well-confined for all delays, with slight changes in peak intensity and mode size with delay. These variations track well with the photodiode trace oscillations in Fig. 2(b) and the expected millisecond-scale thermal evolution of the waveguide core diameter. Based on our measured filament energy deposition and induced density hole depths for single filaments in air [15], we estimate a maximum core-cladding contrast of $\Delta N/N_0 \sim 0.015$ after 500 µs of thermal diffusion, where $\Delta N$ is the air density reduction at the cladding and $N_0$ is the ambient air density in the core. For a waveguide core radius of ~100 µm at 500 µs delay, this gives a $V$ parameter [10,11] of $V \sim 3$, indicating nearly monomode guiding, consistent with the Gaussian-like modes measured.

Figure 3(b) shows the guided mode for several repetition rates at 500 µs delay. The guide confinement improves significantly with repetition rate, consistent with the increase in photodiode signal in Fig. 2. The guide efficiency [10,11] $\eta = (E_g - E_{ug})/(E_{tot} - E_{ug})$ at 500 µs delay is plotted vs. repetition rate in Fig. 3(c), where $E_g$ and $E_{ug}$ are the integrated image signals corresponding to the guided and unguided beam (with and without the waveguide), and $E_{tot}$ is the total beam signal. Efficiency increases with repetition rate, maximizing at ~45% at 1 kHz. In this proof-of-principle experiment, neither the probe injection geometry nor the waveguide cladding were optimized to improve $\eta$ further.

Based on these results, we expect to extend quasi-steady-state air waveguiding to much longer air waveguides [11] driven by wider and higher energy filamenting beams. For such guides, thermal diffusion timescales can be tens of milliseconds [11], likely reducing the required repetition rate below 1 kHz for quasi-continuous operation.

This work is supported by ONR (N00014-17-1-2705 and N00014-20-1- 2233) and AFOSR (FA9550-16-1-0121, FA9550-16-1-0284, FA9550-21-1-0405).